\documentclass[12pt,3p,preprint,sort&compress]{elsarticle}
\pdfoutput=1
\usepackage[colorlinks=true]{hyperref}

\usepackage{amsmath}
\usepackage{amssymb}
\usepackage{mathtools}
\usepackage[utf8]{inputenc}

\renewcommand{\l}{\left(}
\renewcommand{\r}{\right)}

\newcommand{\be}{\begin{equation}}
\newcommand{\ee}{\end{equation}}
\newcommand{\ba}{\begin{align}}
\newcommand{\ea}{\end{align}}
\newcommand{\bg}{\begin{gather}}
\newcommand{\eg}{\end{gather}}
\newcommand{\bseq}{\begin{subequations}}
\newcommand{\eseq}{\end{subequations}}

\newcommand{\Br}{{\rm Br}}
\newcommand{\GeV}{{\rm GeV}}

\begin{document}
\begin{flushright}
	INR-TH-2021-006
\end{flushright}

\title{Double-Hit Signature of Millicharged Particles in 3D segmented neutrino detector} 
\author[inr,mpti]{Dmitry Gorbunov}
\ead{gorby@ms2.inr.ac.ru}
\author[inr]{Igor Krasnov}
\ead{iv.krasnov@physics.msu.ru}
\author[inr,mpti,mephi]{Yury Kudenko}
\ead{kudenko@inr.ru}
\author[inr,mpti,lpnhe]{Sergey Suvorov}
\ead{suvorov@inr.ru}
\address[inr]{Institute for Nuclear Research of Russian Academy of Sciences, 117312 Moscow, Russia}
\address[mpti]{Moscow Institute of Physics and Technology, 141700 Dolgoprudny, Russia}
\address[mephi]{National Research Nuclear University (MEPhI), 115409 Moscow, Russia}
\address[lpnhe]{LPNHE Paris, Sorbonne Universite, Universite Paris Diderot, CNRS/IN2P3, Paris 75252, France}
\begin{abstract}
      We calculate the production of hypothetical millicharged particles (MCPs) of sub-GeV masses by the J-PARC proton beam in the framework of T2K and future T2HK neutrino oscillation experiments. Concentrating on the region of model parameter space, where an MCP can hit the near neutrino detector twice, we adopt this background-free signature to estimate the  sensitivity of T2K  and T2HK experiments to MCPs. We find that  a previously inaccessible in direct searches region of charges 5$\times$$10^{-4}$-$10^{-2}$\,$e$ for MCP masses 0.1-0.5\,GeV can be probed. 
\end{abstract}
\date{}

\maketitle
{\bf 1.} Electric charge quantization remains inexplicable within the Standard Model of particle physics  and may point at some Grand Unified Theory. Therefore it is worth searching for new particles with a fractional electric charge, widely called millicharged particles (MCPs): their observation would imply either a misconception in model building\,\cite{Okun:1983vw} or presence of additional Abelian gauge symmetries in particle physics at high energy\,\cite{Holdom:1985ag}. Even dark matter particles can carry a tiny electric charge, see e.g.\,\cite{Cline:2012is}, with specific consequences for cosmology, and so MCPs can e.g. leave imprints on the anisotropy pattern of the cosmic microwave background\,\cite{Dubovsky:2001tr,Dubovsky:2003yn}. 

All these make MCPs a physically well motivated example of feebly interacting massive particles (FIMPs), which may be light, naturally avoiding detection so far  and requiring a new generation of high intensity frontier experiments\,\cite{Essig:2013lka,Agrawal:2021dbo}. They may be specifically dedicated to searches for new physics projects like SHiP\,\cite{Alekhin:2015byh,Anelli:2015pba}, or may aim at another physics but be capable of performing searches for FIMPs along with working on the main tasks. Among the latter are next generation experiments on neutrino oscillations, and accelerator neutrino experiments, like DUNE and T2HK, cover a wide range of MCP masses\,\cite{Magill:2018tbb}. With huge statistics of protons hitting a target and highly sensitive near detectors primarily devised to control the neutrino fluxes these experiments perfectly meet the criteria of FIMP's hunters.

In this letter, we investigate T2K and T2HK prospects in searches for MCP using the upgraded T2K near detector. Generically neutrino detectors are not suited for the detection of FIMP, since its interaction inside the detector closely mimics that of neutrino. To circumvent this obstacle in a neutrino  experiment, ArgoNeut, Ref.\,\cite{Harnik:2019zee} suggested exploiting the signature of  two subsequent hits inside the detector volume as an MCP candidate. It is feasible for not very tiny electric charge. Simple estimations  show almost no background  from neutrinos produced by the beams. For T2HK oscillation measurements  with $N_{POT}=2.7\times 10^{22}$ protons on target to be collected for   about 10 years of operation and the upgraded T2K near detector designed as described in Ref.\,\cite{Abe:2019whr}, we find this signature very promising. In particular, in models with MCP masses $m_\chi\simeq 0.1-0.5$\,GeV  T2HK will be able to probe previously unattainable region of charges $\epsilon e\simeq 10^{-3}$-$10^{-2}$\,$e$, where $e$ is electron charge.    

It should be mentioned  that a dedicated experiment  to search for MCPs at J-PARC with a sensitivity to  $\epsilon$ of $\sim 10^{-4}$ was proposed in 
Ref~\cite{Kim:2021eix}.  The concept of the detector is based on the idea of a segmented detector comprised of   long scintillator bars with a high photo-electron yield from ionization produced by a charged particle that travels along a bar.

\begin{figure}[!htb]
\begin{center}
\includegraphics[width=0.8\textwidth]{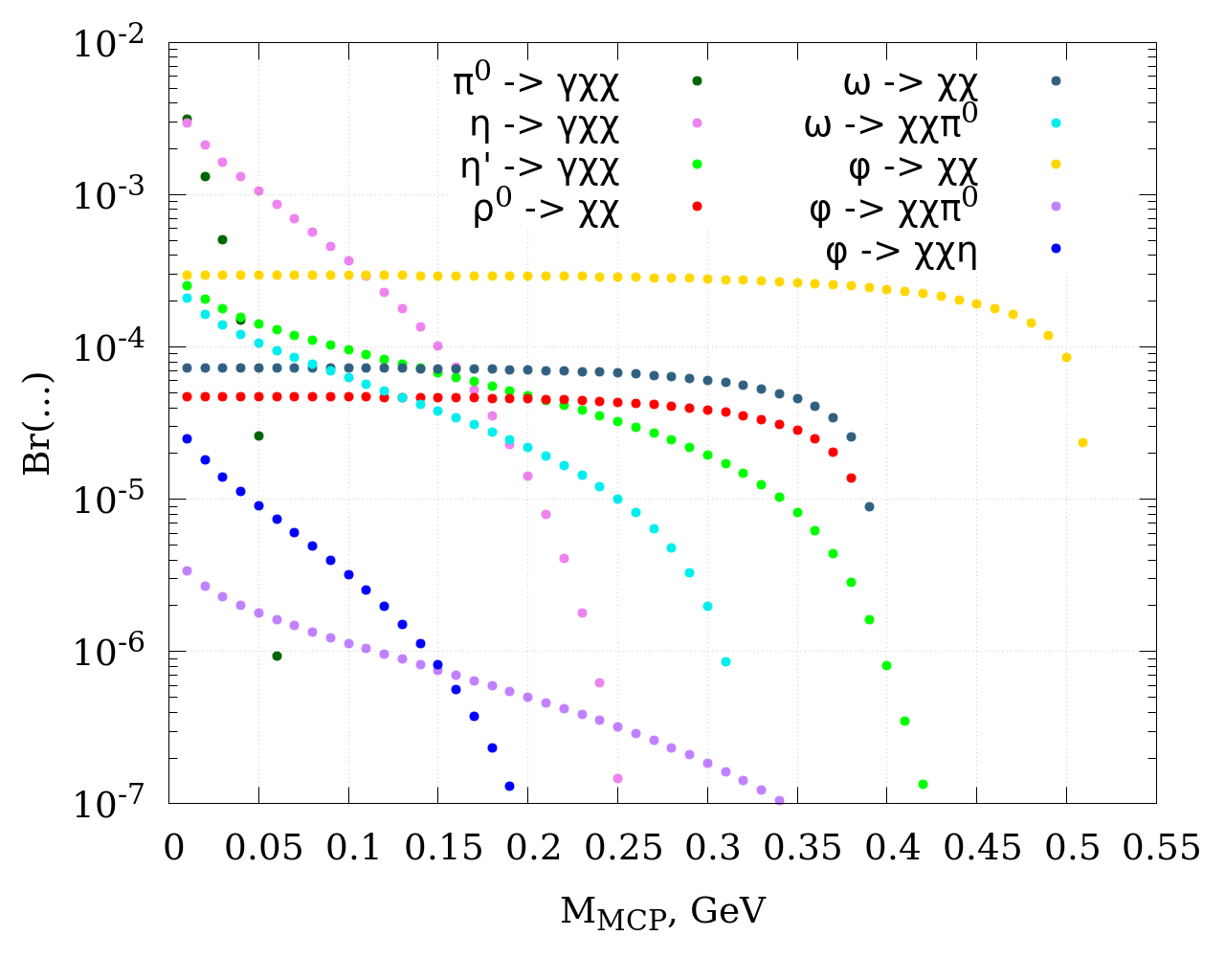}
\caption{Meson branching ratios into MCP for $\epsilon^2 = 1$.}
\label{fig:br}
\end{center}
\end{figure}

{\bf 2.} A pair of MCPs can emerge through a virtual photon in meson decays. This is the main mechanism of the MCP production at JPARC, where a 30\,GeV high intensity proton beam hits the carbon target~\cite{Abe:2011ks} hence generating light mesons. Light vector flavourless mesons $\rho$, $\omega$, $\phi$, can exclusively decay into the MCP pair $\chi\bar\chi$ with branching ratios 
\begin{equation}
\label{2-body}
\Br(V \to \chi \bar{\chi})=\epsilon^2 \cdot \Br(X \to e^+ e^-) \cdot \left(1+2 \frac{m_\chi^2}{M^2_V} \right) \sqrt{1-4\frac{m_\chi^2}{M_V^2}}\,,\;\;\;\;\; V\in\{\rho, \,\omega, \,\phi \}
\end{equation}
obtained by modifying that into muon pair (due to the lepton universality the difference stems from the phase space only)\,\cite{Antipov:1988nk,Ambrosino:2006te}. Pseudoscalar mesons $\pi^0$, $\eta$, $\eta'$ produce a pair of MCPs only in three-body decays, which branching ratios are suppressed with respect to \eqref{2-body} by a phase space factor and additional coupling constants. Vector mesons can decay similarly. Two-body decays are kinematically preferable for a heavier MCP, however, the pseudoscalar mesons are easier to produce in proton collisions, and we account for the three-body processes as well. Their partial decay widths can be derived by generalizing those for electrons and muons in Refs.\,\cite{Ablikim:2015wnx,Arnaldi:2016pzu,Babusci:2014ldz,Anastasi:2016qga}  (see also\,\cite{Kelly:2018brz}), 
\begin{equation}
    \label{3-body}
    \begin{split}
\Br(X \to Y \chi \bar{\chi}) = \epsilon^2 \cdot \Br(X \to Y \gamma) \cdot \frac{2 \alpha}{3\pi} f_{X \to Y} \int_{4 m_\chi^2}^{m_X^2} \frac{dm^2_{\chi\chi}}{m^2_{\chi\chi}} \left(1+2 \frac{m_\chi^2}{m^2_{\chi\chi}} \right) \left(1-4 \frac{m_\chi^2}{m^2_{\chi\chi}} \right)^{\!\!\frac{1}{2}} \\\times
\left(\left(1 +\frac{m^2_{\chi\chi}}{M_X^2-M_Y^2}\right)^2 - 4 \frac{m^2_{\chi\chi} M_X^2}{(M_X^2-M_Y^2)^2} \right)^{\!\!\frac{3}{2}}  \left|F_{XY}(m^2_{\chi\chi})\right|^2,\\
X\to Y\in\{\pi \to \gamma, \eta\to \gamma, \eta'\to \gamma, \omega\to\pi^0, \phi\to\pi^0, \phi\to\eta \}\,, \\
f_{\pi\to \gamma} = f_{\eta\to \gamma} = f_{\eta' \to \gamma} = 1\,,\;\;\; f_{\omega\to\pi^0} = f_{\phi\to\pi^0} = f_{\phi\to\eta} =\frac{1}{2}\,,   
\end{split}
\end{equation}
with $m_{\chi\chi}$ denoting the invariant mass of the MCP pair and form factors taken from Refs.\,\cite{Dzhelyadin:1980kh,Arnaldi:2016pzu,Babusci:2014ldz,Anastasi:2016qga,Ablikim:2015wnx,Beddall:2008zza,Schardt:1980qd} as 
follows
\begin{eqnarray}
|F_{\pi \gamma}(m^2_{\chi\chi})|^2 &=& \left(1+ a_\pi \frac{m^2_{\chi\chi}}{M_\pi^2}\right)^{\!\!2}\!\!,\;\; M_\pi = 0.135 \,\GeV,\;\;  a_\pi =0.11\\
|F_{\eta \gamma}(m^2_{\chi\chi})|^2 &=& \left(1- \frac{m^2_{\chi\chi}}{\Lambda_\eta^2}\right)^{\!\!-2}\!\!\!\!,\;\; \Lambda_\eta =0.72\,\GeV\\
|F_{\eta' \gamma}(m^2_{\chi\chi})|^2 &=& \frac{\Lambda_{\eta'}^2 \left(\Lambda_{\eta'}^2+\gamma_{\eta'}^2\right)}{\left(\Lambda_{\eta'}^2-m^2_{\chi\chi}\right)^2+\Lambda_{\eta'}^2 \gamma_{\eta'}^2},\;\; \Lambda_{\eta'} =0.79\,\GeV,\;\; \gamma_{\eta'} =0.13\,\GeV\\
|F_{\omega \pi^0}(m^2_{\chi\chi})|^2 &=& \left(1- \frac{m^2_{\chi\chi}}{\Lambda_\omega^2}\right)^{\!\!-2}\!\!\!\!, \;\;\Lambda_\omega =0.65\,\GeV\\
|F_{\phi \eta}(m^2_{\chi\chi})|^2 &=& \left(1- \frac{m^2_{\chi\chi}}{\Lambda_{\phi \eta}^2}\right)^{\!\!-2}\!\!\!\!,\;\; \Lambda_{\phi \eta}^{-2} =1.93\,\GeV^{-2}\,\\
|F_{\phi \pi^0}(m^2_{\chi\chi})|^2&=&\left(1- \frac{m^2_{\chi\chi}}{\Lambda_{\phi \pi^0}^2}\right)^{\!\!-2}\!\!\!\!,\;\; \Lambda_{\phi \pi^0} =2.02\,\GeV^{-2}\,.
\end{eqnarray}
The branching ratios entering Eqs.\,\eqref{2-body},\eqref{3-body} are given in \cite{Zyla:2020zbs} as 
\begin{eqnarray*}
\Br(\pi \to \gamma\gamma)= 0.988\,,\;\;\;
\Br(\eta \to \gamma\gamma)= 0.394\,,\;\;\;
\Br(\eta' \to \gamma\gamma)= 0.0221\;\;\;\\
\Br(\rho \to e^+ e^-) = 4.72\times10^{-5},\;\Br(\omega \to e^+ e^-)= 7.28\times10^{-5},\;
\Br(\phi \to e^+ e^-)= 2.95\times10^{-4}
\\
\Br(\omega \to \pi^0 \gamma)= 0.0828\,,\;\;\Br(\phi \to \pi^0 \gamma)= 1.27\times10^{-3}\,,\;\; \Br(\phi \to \eta \gamma)= 0.0131\;\;\;
\end{eqnarray*}
The corresponding partial decay ratios are presented in Fig.\,\ref{fig:br}. 

The light mesons are produced by protons scattering off the target and initiating  hadronic showers. The meson spectra are estimated with the  GEANT4~\cite{Agostinelli:2002hh} package. The choice of the  model of hadronic interactions affects this study, thus various physics lists were considered. The GEANT4 toolkit defines roughly two kinematic regions split with $\approx$ 10 GeV threshold where different models are applied. Above 10 GeV, the most widely used models in HEP are: Quark gluon string model (QGSP) and Fritiof (FTF). At the energy region below 10 GeV we considered BERTini (BERT) and binary cascade (BIC) models. The light meson production was estimated with all the above models and their results were compared. We found nearly no difference ($<$1\%) between the  low-energy BERT and BIC models. While the predictions of QGSP and FTF were quite different. The results of the latter models are summarized in \autoref{tbl:meson_prod}.
\begin{table}[!htb]
    \centering
    \begin{tabular}{c|c|c}
        Meson       & QGSP\underline{{ }}BERT & FTF\underline{{ }}BERT \\
        \hline
        $\pi^0$     & 3.12      & 4.17 \\
        $\eta$      & 0.40      & 0.31 \\
        $\eta'$     & 0.15      & 0.14 \\
        $\rho$      & 0.21      & 0.40 \\
        $\omega$    & 0.12      & 0.27 \\
        $\phi$      & 0.0051      & 0.0051 \\
    \end{tabular}
    \caption{The light meson production per initial 30 GeV proton collision with the  T2K target for different GEANT4 physics lists.}
    \label{tbl:meson_prod}
\end{table}

\begin{figure}[!ht]
    \centering
    \includegraphics[width=\linewidth]{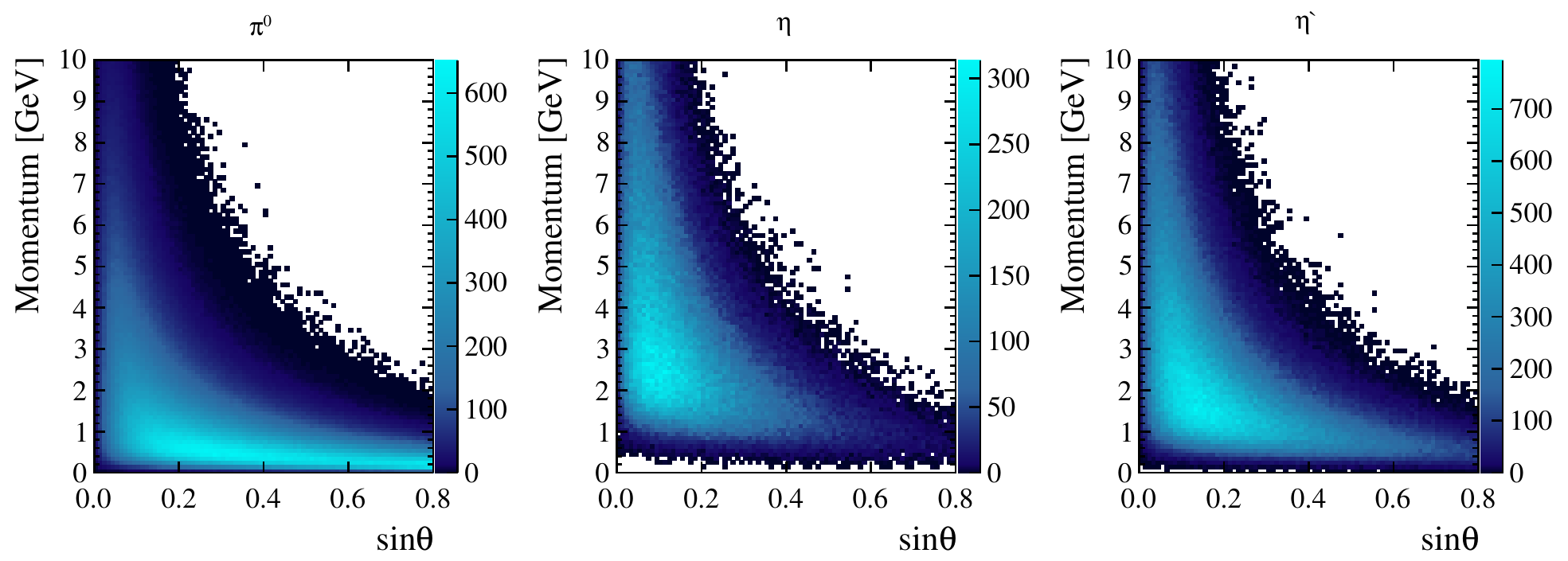}

\centering    
 \includegraphics[width=\linewidth]{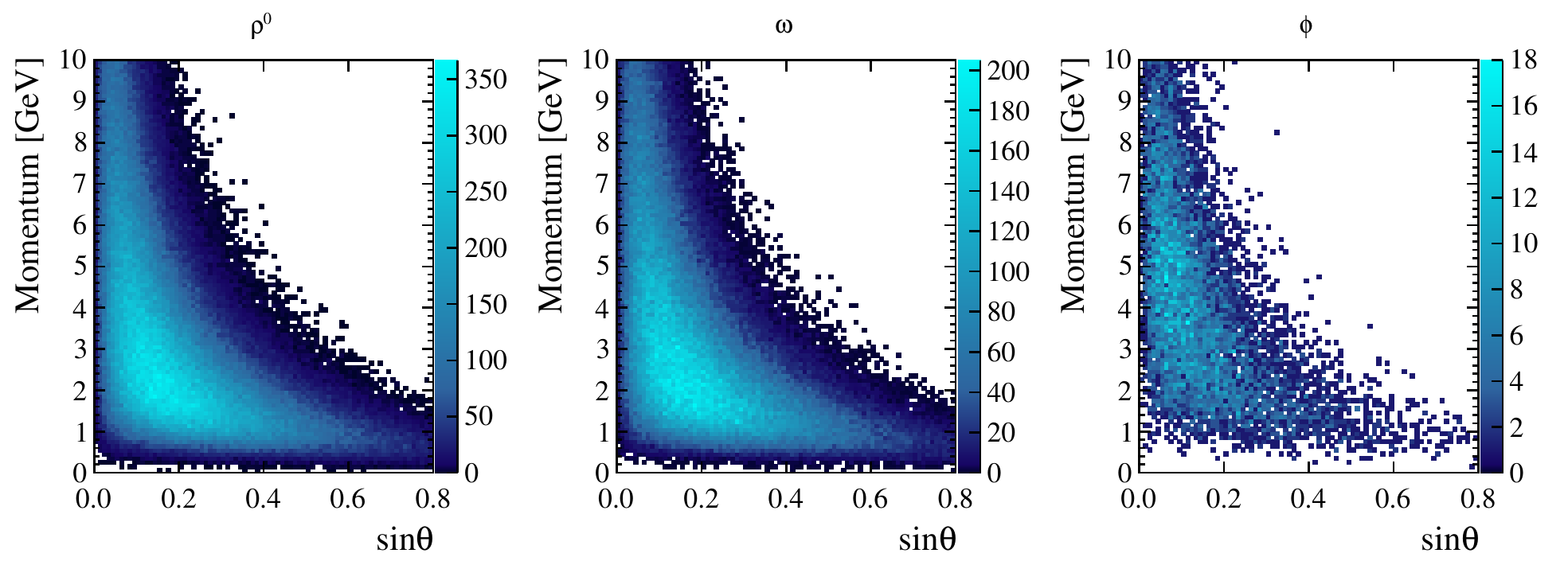}
    \caption{Light meson kinematic distributions over 3-momentum and angle with respect to the J-PARC proton beam.}
    \label{fig:meson_kinem}
\end{figure}

The QGSP\underline{{ }}BERT physics list was considered more reliable as it provides better agreement with known T2K $\pi^0$ production and we use it in what follows. However, we found no experimental measurements of the production rates of other light mesons, thus the number of initial mesons is a possible source of  uncertainties in the current study. The largest difference between the models was observed for $\rho$ and $\omega$ yields. The kinematic distributions of the produced mesons are shown in Fig.\,\eqref{fig:meson_kinem}.

We performed simulations for $N_{sim}=2\times 10^6$ protons on target (POT) which reveal the following (approximate) numbers of produced light mesons participating in the MCP phenomenology:  
\begin{eqnarray*}
&&N_\pi = 6.24 \times 10^6\,,\;\;N_\eta  =  7.94 \times 10^5\,,\;\;
N_{\eta'}  =  2.96 \times 10^5\,,\\
&&N_\rho  =  4.16 \times 10^5\,,\;\;N_{\omega}  =  2.32 \times 10^5\,,\;\;
N_{\phi}  =  1.01 \times 10^4\,.
\end{eqnarray*}
Directions of outgoing MCPs are obtained 1) adopting isotropic distribution in the rest frame of decaying mesons in case of two-body decays, and for the three-body decays we choose the invariant mass of MCPs (and thus the energy of the third particle) randomly in accordance with distribution  \eqref{3-body} and assigning the third particle's momentum a random direction in the rest frame we restore the momenta of MCPs accordingly, and 2) performing the Lorentz transformation back to the laboratory system with a help of boost along the decaying meson 3-momentum.  To be detected, the produced MCP must   make the selected signature of two subsequent hits inside the detector volume, and so we require for the trajectory of observable MCP to pass through the T2K near detector, which is placed at a distance of $d=280$\,m from the target and at 2.5$^\circ$ off the proton beam axis. 

\begin{figure}[!htb]
\begin{center}
\includegraphics[width=0.8\linewidth]{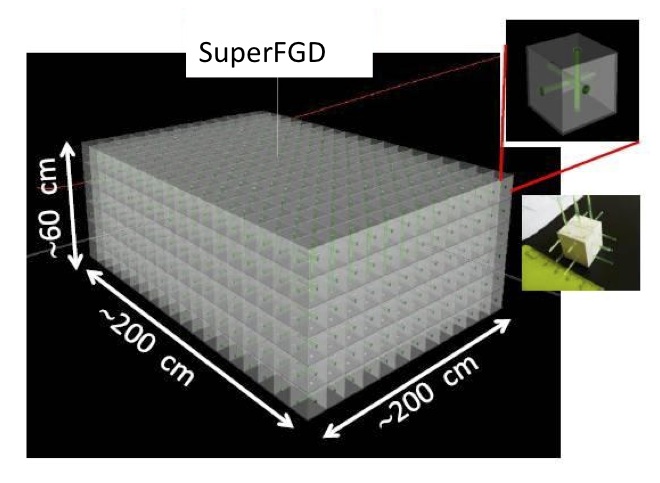}
\caption{3D view of    the SuperFGD structure. Also shown are  cubes of  1$\times$1$\times$1\,cm$^3$ with 3 orthogonal wave-length shifting fibers inserted into holes.}
\label{fig:superfgd}
\end{center}
\end{figure}

{\bf 3.} For the MCP detection we consider  the new neutrino detector SuperFGD~\cite{Sgalaberna:2017khy} which will be installed  inside the off-axis detector complex ND280. The main purpose of this detector is to  reduce the systematic uncertainties in the prediction of a total number of signal neutrino events in the far T2K detector Super-Kamiokande, in presence of oscillations~\cite{Abe:2019whr}. SuperFGD begins data taking within the T2K program, operating with Super-Kamiokande~\cite{Abe:2018uyc}, and then will be used for measurement of CP asymmetry in neutrino oscillations with the  Hyper-Kamiokande detector (T2HK program). 
The highly granular scintillator detector SuperFGD of a mass of about 2 tons  is comprised of  $\sim 2\times 10^6$ small scintillator cubes of 1 cm side, each read out with WLS fibers in the three orthogonal directions coupled to compact photosensors,  Multi-Pixel Photon Counters (MPPCs), as shown in Fig.\,\ref{fig:superfgd}.

\begin{figure}[!htb]
\begin{center}
\includegraphics[width=0.8\textwidth]{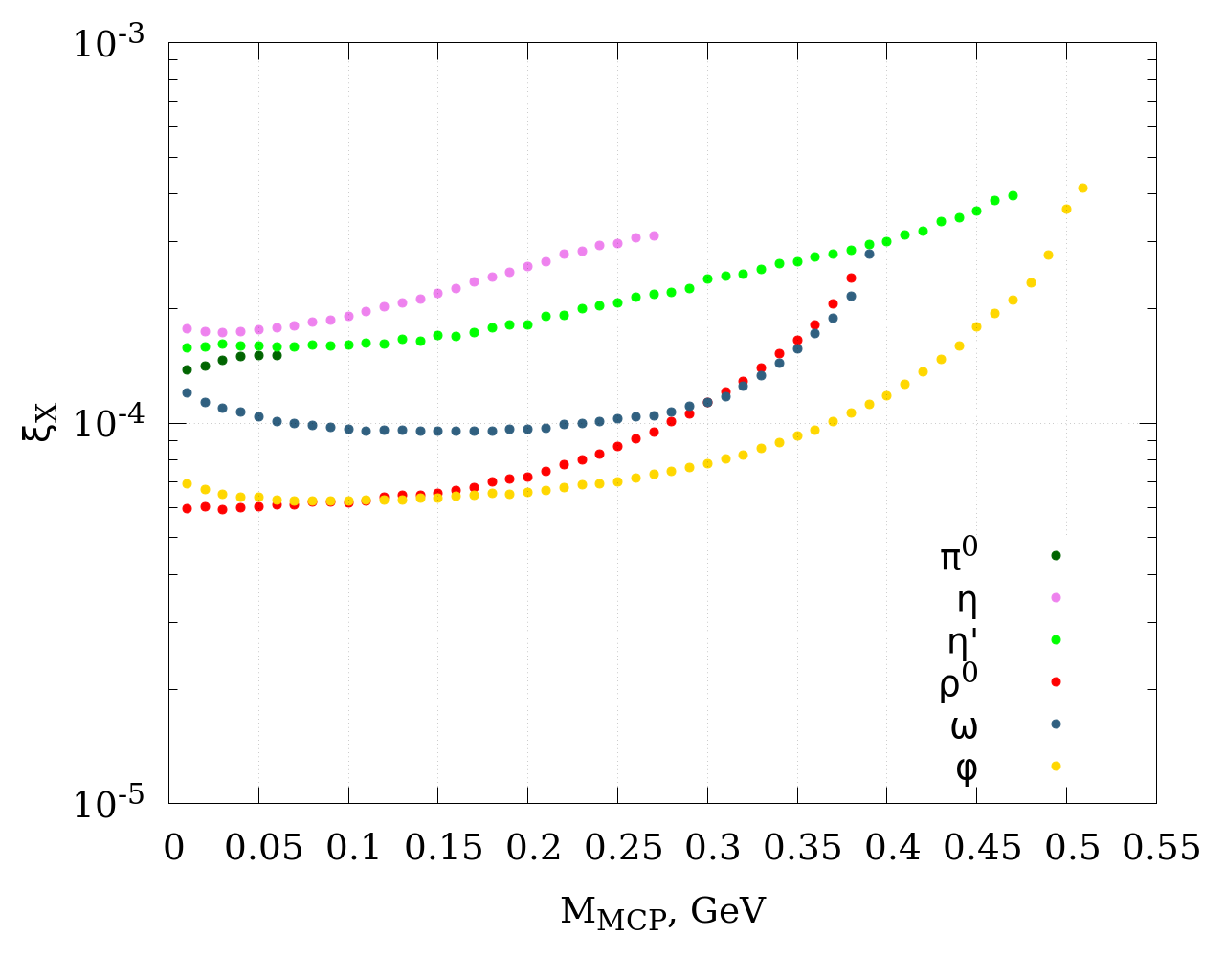}
\caption{The geometrical factor $\xi_{X,i}$ for MCP as a function of its mass  for each  parent meson.} 
\label{fig:xi}
\end{center}
\end{figure}

\begin{figure}[!htb]
\begin{center}
\includegraphics[width=0.8\linewidth]{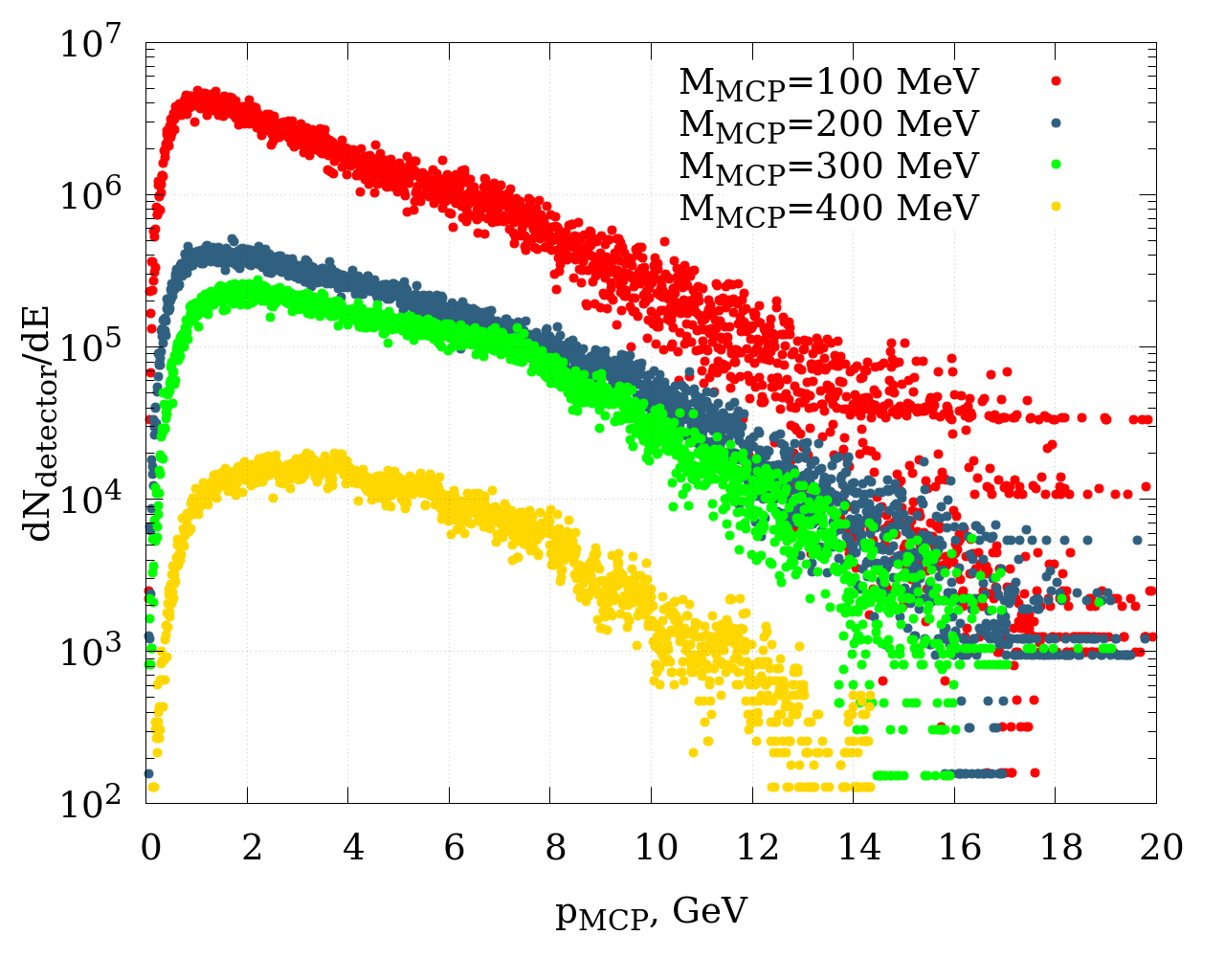}
\caption{The spectra of MCPs that reach the SuperFGD for a set of masses and $\epsilon = 10^{-3}$. Integrating the spectrum over $E$ reveals the total number of MCPs that  pass through the detector for $N_{POT}=3.2\times 10^{22}$: $\,2.01\times10^9$ for $M_{MCP}=100$\,MeV, $2.63\times10^8$ for $M_{MCP}=200$\,MeV, $1.58\times10^8$ for $M_{MCP}=300$\,MeV, $1.15\times10^7$ for $M_{MCP}=400$\,MeV.}
\label{fig:spectra}
\end{center}
\end{figure}

SuperFGD will serve  as an active neutrino target and a $4\pi$ detector of charged particles from neutrino interactions.  The size of SuperFGD  is about  $0.56\times1.92\times1.84$\,m$^3$. A small angle 2.5${}^\circ$ with respect to the neutrino beam doesn't cause a strong reduction of the MCP flux. The main acceptance limitation comes from the detector front surface area. The detector is placed so that its front side with respect to the beam has a size of 1.92$\times$0.56\,$\text{m}^2$ and the 1.84\,m side is oriented along the beam direction. We define a geometrical factor $\xi_{X,i}$ as the fraction of simulated MCP trajectories entering SuperFGD. These factors are calculated for each MCP production mode, the results are summarized for each parent meson in Fig.\,\ref{fig:xi}. 
The corresponding numbers and spectra of MCPs that reach the detector are presented in Fig.\,\ref{fig:spectra}. 
One can check along the lines of Ref.\,\cite{Harnik:2019zee} that for the reference value of $\epsilon=10^{-3}$ the energy loss and trajectory deflections due to MCP multiple scattering in soil on the way of  $\sim 200$\,m to the detector  are negligible. 

Prototypes of SuperFGD were tested in a charged beam at  CERN and  showed a  very good performance~\cite{Mineev:2019dpe,Blondel:2020hml}. Light yields per minimum ionizing particle (MIP) of 50-60 photoelectrons (p.e.) from individual cubes and  from a single WLS fiber were obtained. The sum of signals from 3 WLS fibers gives  the total light yield of 150-180 p.e. per MIP  for a single cube. Very good timing was also obtained in the beam tests. The time resolution of 
$\sigma \sim 1$\,ns  for an individual cube for the light yield corresponding to about  2 MeV energy deposited in this cube and   measured by one WLS fiber. For 3 fiber readout the time resolution is expected to be about 0.6 ns  in this case. For a larger than 2 MeV energy deposit in SuperFGD (more than 1 cube is fired) the time resolution should be better than 0.5 ns. This parameter is important for the suppression of the neutron background produced by the neutrino  beam  as discussed below.   

SuperFGD will be equipped with Hamamatsu MPPCs  S13360-1325 which have unique features: a very low  dark rate of 60-70 kHz  and 0.5 kHz at the threshold of 0.5 p.e. and 1.5 p.e., respectively, and a low cross-talk of about 1\% \cite{Blondel:2020hml}. The detection signature of MCPs in the SuperFGD detector is elastic scattering off atomic electrons, which results in knock-on $\delta$-electrons above the detection threshold providing a detectable signal.  Assuming  that parameters of SuperFGD  will be close to those obtained in the beam tests, one can expect  that the  energy  of about 100 keV deposited by a recoil electron produces  the light yield of 3.1-3.6 p.e. per a WLS fiber. Given this result,  the efficiency to detect a 100 keV electron  in one WLS fiber is estimated to be  $\geq 82$\% for the threshold of 1.5 p.e. 
Since about 99\% of recoil electrons have energies from 0.1 MeV to 10 MeV,   the total efficiency    to detect recoil electrons by one WLS fiber is estimated to be about 98\%.     As a result, one can expect to reach the MCP detection efficiency in an individual cube of  about 96\%  for a coincidence of signals from two WLS fibers  for the energy  threshold of $E_r^{min}=100\,\text{keV}$.     The  idea to detect MCPs  lays in using two separated hits from an MCP aligned with   the upstream meson production target. This method,  proposed  in Ref.~\cite{Harnik:2019zee} and used  for a search of MCPs in the ArgoNeuT experiment~\cite{Acciarri:2019jly}, allows us to achieve a good background rejection, as seen below.    

{\bf 4.} It is important to investigate a  two-hit background in SuperFGD from several sources. Signals  caused by random electronics noise due to the MPPC dark rate  can mimic the MCP signal. Assuming  the mentioned above MPPC dark rate of 0.5 kHz for the threshold of 1.5 p.e., the time window of 30\,ns for each readout channel, and the two-fiber readout (both fibers are perpendicular to the beam direction) one can obtain the counting rate   of about $1.5\times 10^{-2}$\,s$^{-1}$ due to coincidence of noise signals from two MPPCs. The total  dark rate from all $2\times 10^6$ SuperFGD  cubes will be about $3\times 10^4$\,s$^{-1}$. Since the  accidental  hit events will be uniformly distributed in SuperFGD volume they will only rarely align with the  meson production target. To estimate the number of two electronic noise  hits (two distant cubes in a line with the production target provides signals above the threshold of 1.5 p.e.) we assume that the second hit occurs inside a 5$\times$ 5 cluster in a column of about 100 cubes  that forms a straight line with the first cluster and the target. The rate of such coincidence  is estimated to be  $\leq 4\times 10^{-12}$\,s$^{-1}$. Taking into account  the beam structure   with the spill width of 5\,$\mu$s and the  repetition period of 2.48\,s  for T2K (1.16\,s for T2HK)~\cite{Abe:2018uyc},  accidentals in a cube  due to the dark rate of MPPCs  are expected to be  suppressed by a factor of $2\times 10^{-6}$ for T2K and  $< 5\times 10^{-6}$ for T2HK.  Assuming $2.7\times 10^{22}$ POT for Hyper-Kamiokande that corresponds to about $10^8$\,s of data taking, the expected  total number of background events due to the random electronics noise is  $\leq 10^{-4}$.    

Another source of accidental background can be  the coincidence between the signal from  a cube where   MPPCs dark rate    mimics the  MCP signal  (the first hit) and the real MCP signal from its interaction  in SuperFGD (the second hit).  Assuming the interaction  length of MCP ($\epsilon = 10^{-3}$) is about $1.3\times 10^6$\,cm$^2$,  for a cluster of $5\times 5$ cubes which is on  a straight line with the first cluster and the target,  one can obtain the number of such coincidences is about $2.5 \times 10^{-2}$ for running time of $10^8$\,s.

The vast majority of neutrino induced events, for example, double-hit events from muon (electron) and neutron in the case of $\nu_{\mu}$ charge current quasi elastic scattering (CCQE)  will provide signals with   a large number of cubes fired.  By implementing the requirement that the  track length  should be $\leq 5$ cubes that corresponds to  the electron energy deposit of $\leq 10$\,MeV  these neutrino induced backgrounds can be significantly reduced.   

The neutral current reactions in SuperFGD 
    \begin{equation}
    \label{nc-1}
 \nu + {}^{12}\text{C} \to \nu^{\prime} + {}^{11}\text{C}^{\star} + n 
\end{equation}
     and 
 \begin{equation} 
 \label{nc-2}
     \nu + {}^{12}\text{C} \to \nu^{\prime} + {}^{11}\text{C} + n
   \end{equation}  
 are the most serious sources of background.  There are no  measurements of these cross sections on $^{12}{\rm C}$  at the T2K neutrino energies, but for the  background estimation,  we can use the value of   the  neutral-current elastic-like cross section   on oxygen $\sim 10^{-38}$ cm$^2$   measured by T2K~\cite{Abe:2019cpx} via detecting nuclear de-excitation $\gamma$-rays at Super-Kamiokande. 
   
The excited $^{11}{\rm C}^{\star}$ emits $\gamma$-rays promptly and relaxes to the ground state. Assuming that the detection efficiency of $\gamma$-rays produced from the de-excitation of $^{12}{\rm C}^{\star}$ is 100\%, one can find that  about $3\times 10^5$ such events will be detected in SuperFGD for $2.7\times 10^{22}$ POT. The neutron from the reaction~(\ref{nc-1})  can mimic the MCP signal if its hit is in coincidence with the neutrino interaction vertex time, both neutrino and neutron  vertexes align with the  target, and neutron deposits the energy of $\leq 15$ MeV in its interaction because less than 1\% of  knock-on $\delta$-electrons  from the MCP interaction have the energy exceeding 15 MeV.   Taking into account the   time-of-flight between the first signal (de-excitation of $^{11}{\rm C}^{\star}$) and the second one from the scattered neutron, these background events will be suppressed using excellent time resolution of $\sigma \sim 0.2-0.3$\,ns. It can be obtained for events in  which several cubes are fired that corresponds to energies of a few MeV. 
Neutrons with the  kinetic energy below 300\,MeV (about 90\% of all events) are estimated to be rejected  by a factor of  $10^4$ with the  time-of-flight method. Neutrons in the energy range 300-600\,MeV can be suppressed by three orders of magnitude. A small amount of neutrons with energies exceeding 600\,MeV ($< 10^{-2}$) will be rejected by about 20 times.  The  requirement for  protons from $(n,p)$ scattering to  have the kinetic energy  of $\leq 15$\,MeV gives an additional suppression factor of about 10. The cross section of the  reaction~(\ref{nc-2})  with $^{11}{\rm C}$ in the ground state is several times smaller than the cross section of (\ref{nc-1}). To mimic the MCP event, the neutron from this reaction should interact two times in the SuperFGD and pass through all selection criteria.  Using the  approach applied for the reduction of the background from the reaction~(\ref{nc-1}), the background from the reaction~(\ref{nc-2}) is expected to be suppressed to a much lower level.  
   
Neutrons produced by the beam neutrinos in the sand around the ND280 pit  and in the ND280 magnet interact many times and  are slowed down during the propagation. A large fraction of reached SuperFGD neutrons  are delayed with respect to the neutrino interaction time and  produce signals between the beam bunches  and microbunches (8 microbunches each of about 50\,ns  separated by 700\,ns fill a beam bunch of 5\,$\mu$s). Events in  which neutrons   coincide with the beam spill and interact 2 times in  SuperFGD will be suppressed by the time-of-flight method and the required  alignment of two hits with  the target. 
    
In total,  we can expect  less than 0.1 event contribution to the MCP  background  from the neutron current interactions in SuperFGD.    These estimations of  background rates  are used  to calculate the expected sensitivity of SuperFGD to MCPs. Eventually, the background will be determined from the data accumulated with the neutrino beam.

{\bf 5.} Upon entering the detector, an MCP can scatter off the material electrons.  The recoil electrons can be observed, if the recoil energy is above a certain threshold $E_r^{min}$, and so the interesting cross section $\sigma(E_r^{min})$ depends on the energy carrying away by the electron. In the limit of a relativistic MCP the mean free path of MCP inside material with electron  density $Zn_{det}$ reads\,\cite{Magill:2018tbb}
\begin{equation}
\lambda =  \frac{1}{Z n_{det} \sigma(E_r^{min})}=\epsilon^{-2}\frac{m_e E_r^{min}}{2\pi \alpha^2 Z n_{det}}\,.
\end{equation}
For  SuperFGD (made up of carbon-based scintillator) we have $Z=7$, and the matter density is $\rho_{ND} = 1.0 \frac{\text{g}}{\text{cm}^3}$   with molar mass $m_a(CH) = 13.0 \frac{\text{g}}{\text{mol}}$~\cite{Blondel:2020hml}, and hence $n_{det} = \frac{\rho_{ND}}{m_a(CH)} = 4.64\times 10^{22}$\,cm$^{-3}$. 
Consequently, for the MCP mean free path one obtains
\[
\lambda\approx1.2\times 10^{4}\times\l\frac{10^{-3}}{\epsilon}\r^2 
\times \l\frac{E_r^{min}}{100\,\text{keV}}\r\text{m}\,.
\]
normalized to the expected SuperFGD threshold of the electron recoil energy $E_r^{min}=100$\,keV. 

In this study, we utilize a double-hit signature: the MCP has to scatter twice inside the detector, each time transferring to electron the energy above the threshold  $E_r^{min}$. 
The possibility of 2 consecutive hits, each observed with efficiency $\xi$,  is\,\cite{Magill:2018tbb}:
\begin{equation}
P_{2h} = \frac{1}{2} \left(\xi\,\frac{L}{\lambda}\right)^2 =  \frac{1}{2} \left(\frac{0.96\times\l\frac{\xi}{0.96}\r\,1.84\,\text{m}}{\left(\frac{10^{-3}}{\epsilon}\right)^2 \left(\frac{E_r^{min}}{100\,\text{keV}}\right) 12\,\text{km}}\right)^2 \approx 1.1\times 10^{-8}\times \l\frac{\epsilon}{10^{-3}}\r^4.
\end{equation}
The detection efficiency $\xi$ of each MCP hit  with the electron energy $\geq 100$  keV is estimated to be about 96\%.
Remarkably, the probability to produce the chosen signature does not depend on the MCP production channel.

\begin{figure}[!htb]
\begin{center}
\includegraphics[width=0.8\linewidth]{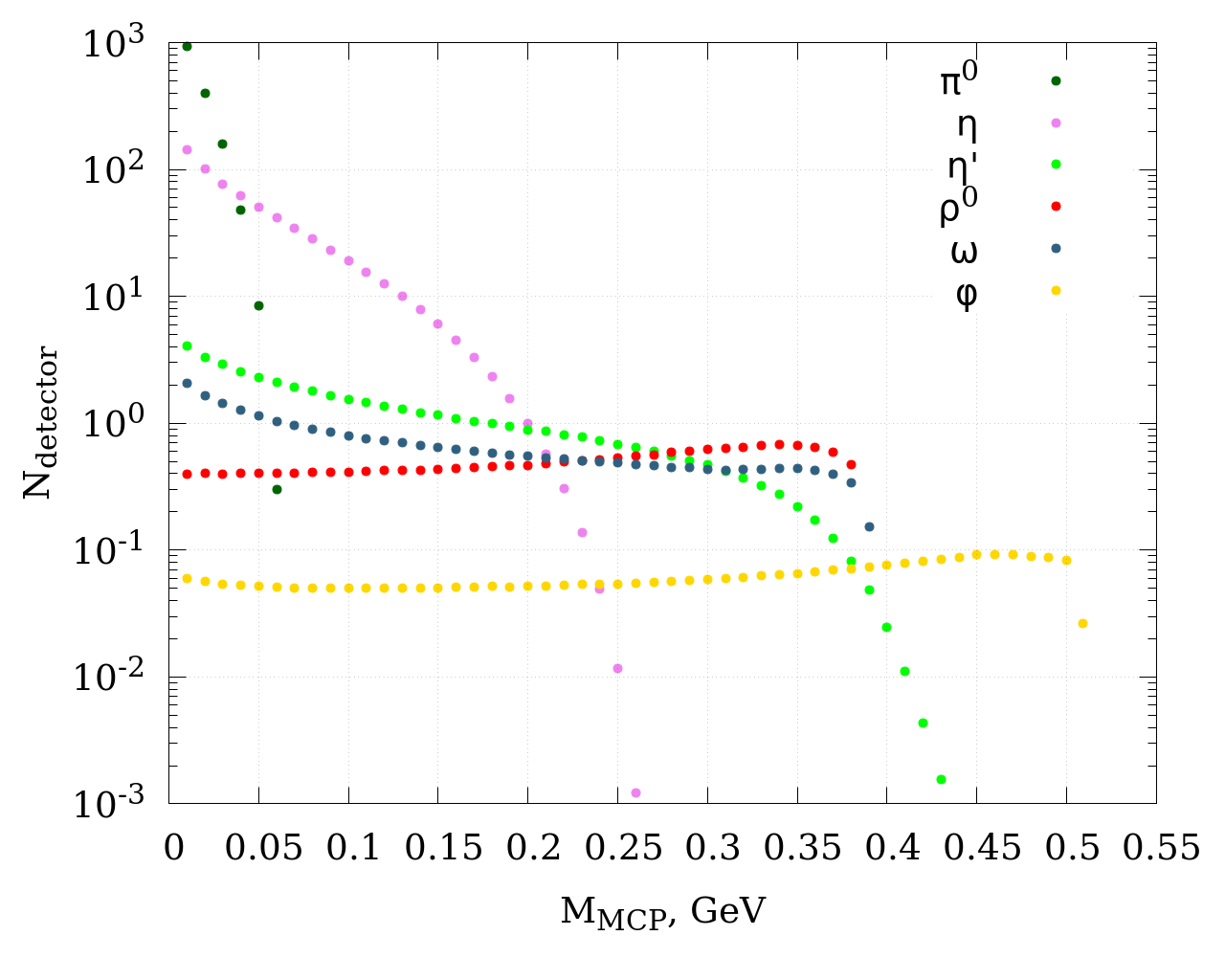}
\caption{The simulated number of the double-hit MCP events in SuperFGD for various parent mesons and $\epsilon = 10^{-3}$.}
\label{fig:detector}
\end{center}
\end{figure}

\begin{figure}[!htb]
\begin{center}
\includegraphics[width=0.8\linewidth]{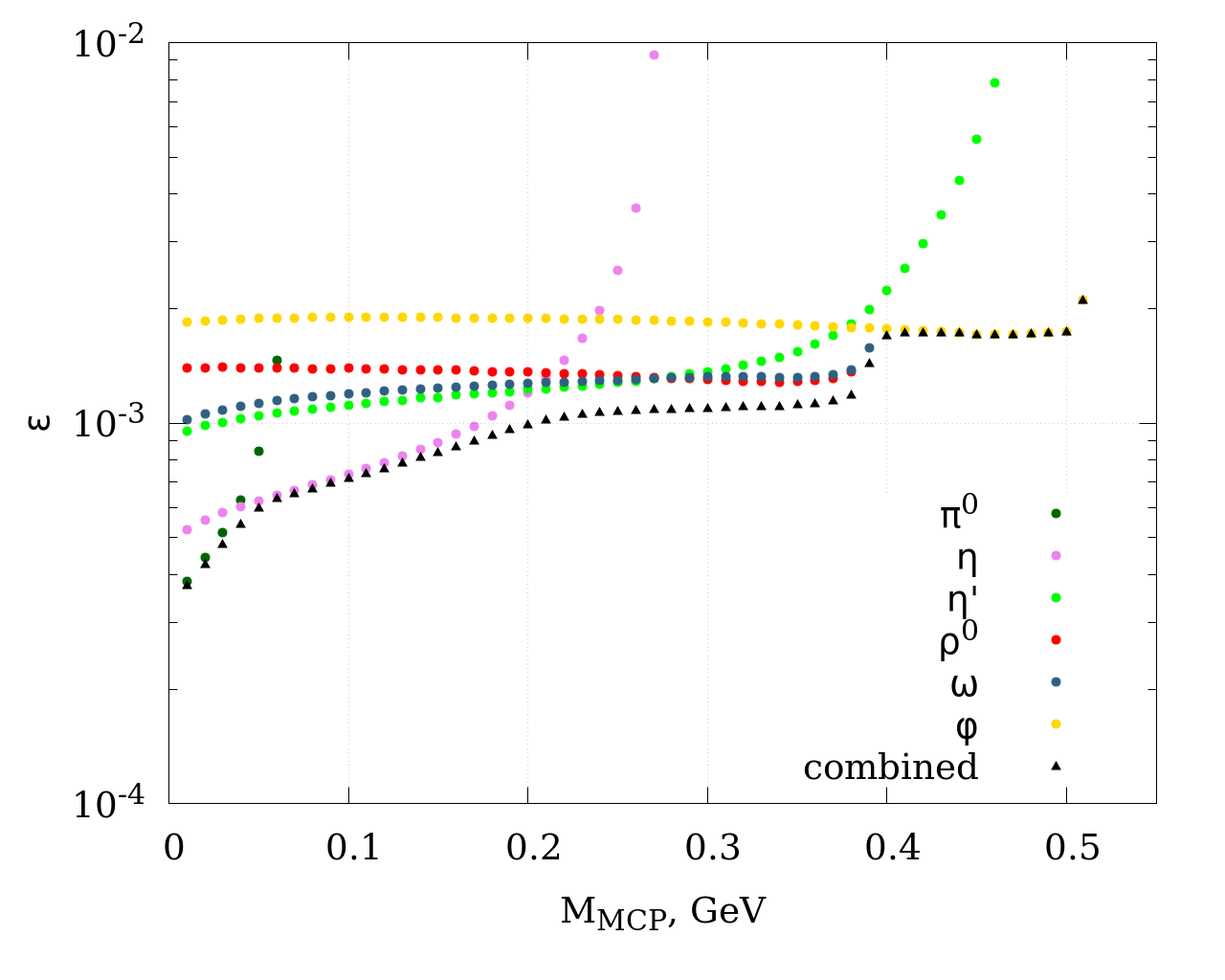}
\caption{The expected exclusion regions (above the dots) of $\epsilon$ in case of 3 signal events and no background.}
\label{fig:e2}
\end{center}
\end{figure}

{\bf 6.} At this stage we can sum up contributions of various production modes to the number of signal events $N_S$. Our estimate of $N_S$ is based on the described above GEANT4 simulation of the production of the meson of type $X$, its branching ratios in a particular decay mode $\Br_i(X\to\dots)$, and calculation of the corresponding geometrical factor $\xi_{X,i}$, as follows 
\[
N_S=N_{POT}\times \sum_X\frac{N_X}{N_{sim}}\times \sum _i \Br_i(X\to\dots)\times\xi_{X,i}\times P_{2h}\,.
\]
In Fig.\,\ref{fig:detector} 
we plot for each parent meson the number of expected events for MCP reference model with  $\epsilon=10^{-3}$. Since the signature we adopt  depends on neither energy nor mass of the MCP, we can estimate the total number of expected signal events in SuperFGD for each MCP mass simply summing over all the production channels. Requiring this number to be smaller than 3 we estimate the T2HK sensitivity (at 95\% CL) to the MCP charge: the region above the black dots in Fig.\,\ref{fig:e2}  
will be excluded after 10 years of data taking (in case of no signal). We draw the exclusion curves for each parent meson to illustrate for which MCP mass they give a dominating contribution.   

\begin{figure}[!htb]
\begin{center}
\includegraphics[width=0.8\linewidth]{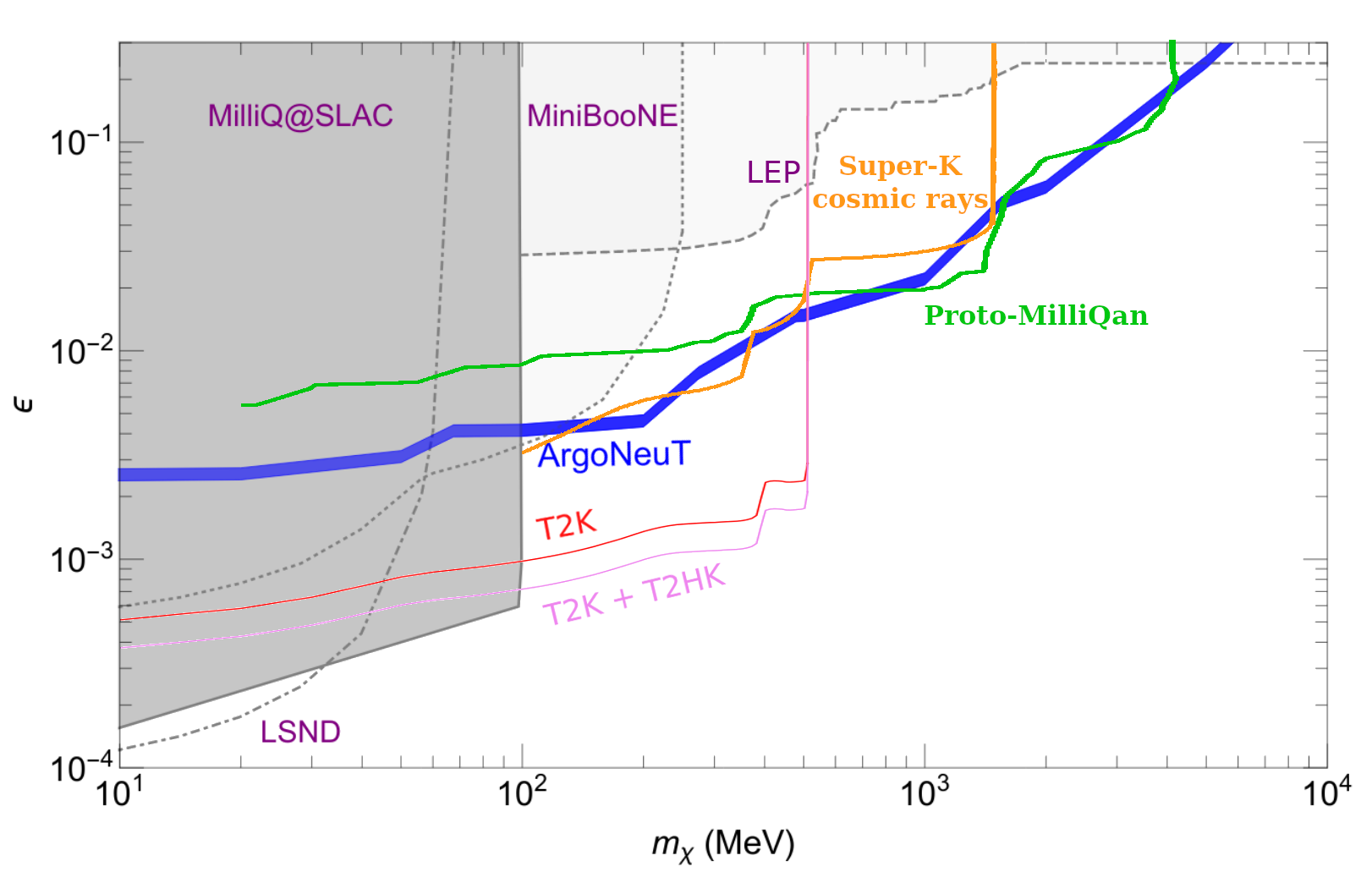}
\caption{The expected sensitivity of T2K and T2HK to models with millicharged particles. The regions indicated by MilliQ@SLAC\,\cite{Prinz:1998ua} and ArgoNeut\,\cite{Acciarri:2019jly}, recent LHC results (Proto-MilliQan) \cite{Ball:2020dnx}  are excluded by the corresponding collaborations. There are also regions disfavored after reanalysis of published results by MiniBooNE, LSND,\,\cite{Magill:2018tbb} and LEP, see e.g.\,\cite{Davidson:2000hf},  and analysis of the Super-K results with cosmic rays producing the MCPs \cite{Plestid:2020kdm}.}
\label{fig:results}
\end{center}
\end{figure}

{\bf 7.} 
To conclude, in this paper we propose to use the double-hit signature of hypothetical millicharged particles at presently under construction SuperFGD Near Detector of T2K-T2HK long-base line neutrino experiment. We evaluate the production of MCP particles and trace their trajectory passing through the detector. We argue that the signature is background free for the expected T2HK statistics of protons on target. Since SuperFGD will operate for  a few years within the T2K program and then switch to the T2HK program we calculate the expected number of events for both stages of operation, assuming $0.5\times 10^{22}$\,POT and $2.7\times 10^{22}$\,POT respectively. Assuming no signal events to be observed and exploiting the Poisson statistics we present in Fig.\,\ref{fig:results} the expected sensitivities (95\% CL exclusion regions) for T2K  and for the sum of both T2K and T2HK experiments.    
There are also limits placed by dedicated searches of MiiliQ@SLAC\,\cite{Prinz:1998ua} and ArgoNeuT\,\cite{Acciarri:2019jly} and specific analyses of other accelerator data,  including recent LHC results \cite{Ball:2020dnx}.   
 Also, results of the Super-K experiment were used in Ref.\,\cite{Plestid:2020kdm} to constrain the model parameter space, considering possible MCP production by cosmic rays. While it is similar to the ArgoNeuT results, there are significant uncertainties associated with the description of the propagation and interaction of the particles in the atmosphere and a more detailed analysis of the signal signature and associated background conditions would be appropriate.  
The suggested searches at T2K and T2HK will investigate untouched by those searches region of masses 100-500\,MeV and charges as low as $\sim 5\times 10^{-4}$ of the electron charge. Recently it was argued\,\cite{Marocco:2020dqu} that a  noticeable part of this region is disfavored from the results of the BEBC experiment operated in the 1980s at CERN. 

The sensitivity is obtained assuming the negligible background, for chosen cuts on the recoil electron energy and the expected detection efficiency. All that should be checked after the commissioning stage of SuperFGD, which may allow one to   optimize our set of cuts and efficiencies and further refine the obtained sensitivity. Special attention should be paid to the background from non-relativistic neutrons since the signal recoil events are in a rather low,  sub-MeV energy range.    

There are also some uncertainties on the theoretical side of calculations that are associated with the observed dependence of GEANT4 simulations of the meson production on the chosen QCD model. 

However, since the number of signal events scales as the sixth power of the MCP charge, the overall uncertainty of the presented in Fig.\,\ref{fig:results} sensitivity is small, and our predictions are robust. At the same time, this strong dependence on the MCP charge makes any further improvement in the presented techniques rather fruitless. To investigate models with a light particle of smaller electric charge one must rely on other signatures with the number of signal events involving lower powers of MCP charges, the {\it potentially} most promising is just missing energy, so the number of events is proportional to the squared charge only, like e.g. in ongoing NA64\,\cite{Gninenko:2018ter}.

\vskip 0.5cm

We thank C.K.Jung, M.Khabibullin, T.Matsubara for valuable discussions and M.\,Kirsanov, N.\,Starkov for clarification on GEANT4 packages. This
work is supported in the framework of the State project “Science” by the Ministry of Science
and Higher Education of the Russian Federation under the contract 075-15-2020-778.
\bibliographystyle{utphys}
\bibliography{refs}
\end{document}